\documentclass[aps,twocolumn,longbibliography]{revtex4-1}
\pdfoutput=1

\usepackage[lmargin=2cm,rmargin=2cm,tmargin=2cm,bmargin=2cm]{geometry}
\usepackage{amsmath,amssymb,mathtools,bm,color,graphicx}

\newcommand{\pv}{\operatorname{P}}

\begin{document}

\title{Probability backflow for correlated quantum states}

\author{Arseni Goussev}

\affiliation{School of Mathematics and Physics, University of Portsmouth, Portsmouth PO1 3HF, United Kingdom}

\date{\today}

\begin{abstract}
	In its original formulation, quantum backflow (QB) is an interference effect that manifests itself as a negative probability transfer for free-particle states comprised of plane waves with only positive momenta. Quantum reentry (QR) is another interference effect in which a wave packet expanding from a spatial region of its initial confinement partially returns to the region in the absence of any external forces. Here we show that both QB and QR are special cases of a more general classically-forbidden probability flow for quantum states with certain position-momentum correlations. We further demonstrate that it is possible to construct correlated quantum states for which the amount of probability transferred in the ``wrong'' (classically impossible) direction exceeds the least upper bound on the corresponding probability transfer in the QB and QR problems, known as the Bracken-Melloy constant.
\end{abstract}

\maketitle

\section{Introduction}

Quantum backflow (QB) is a quantum-mechanical interference effect with no counterpart in classical mechanics. It manifests itself as a flow of probability density in the direction opposite to the direction of the momentum of a quantum particle. For a more concrete definition of QB, let us consider a free particle travelling along the $x$-axis. Suppose that it is know {\it with certainty} that the momentum $p$ of the particle is positive, i.e. $p > 0$. Let $P_<(t)$ denote the probability that, at time $t$, the particle is located to the left of the origin, i.e. at $x < 0$. The quantity of interest is
\begin{equation}
	\Delta(\tau, T) = P_<(\tau + T) - P_<(\tau) \,.
\label{Delta_def}
\end{equation}
where $\tau \in \mathbb{R}$ and $T > 0$. The physical meaning of $\Delta(\tau, T)$ is the amount of probability that has been transported from the right of the origin, $x > 0$, to the left of the origin, $x < 0$, during the time interval $\tau < t < \tau + T$. If the motion of the particle was governed by the laws of classical mechanics, it would be impossible for $\Delta(\tau, T)$ to exceed zero: in classical mechanics, the direction of the probability flow is the same as that of the particle momentum. In quantum mechanics however there exist wave packets, composed entirely of plane waves with positive momenta, for which $\Delta(\tau, T) > 0$; this is the QB effect.

The first argument for the existence of the QB effect was made in Refs.~\cite{All69time-c, Kij74time}. There it was pointed out that a (non-normalizable) linear combination of two plane waves with positive momenta may generate a negative probability current. QB for normalized wave packets was first analysed by Bracken and Melloy~\cite{BM94Probability}. In particular, they showed that the supremum of the right-to-left probability transfer -- $\sup \Delta(\tau, T)$ where the supremum is taken over all normalizable states comprised of plane waves with positive momenta -- coincides with the supremum $\lambda_{\sup}$ of the eigenvalue spectrum in the following integral eigenproblem:
\begin{equation}
	-\frac{1}{\pi} \int_{0}^{\infty} du' \, \frac{\sin \left( u^2 - {u'}^2 \right)}{u - u'} \psi(u') = \lambda \psi(u) \,,
\label{int_eig_prob}
\end{equation}
where $\psi(x)$ belongs to the class of square-integrable functions defined on $x > 0$. That is, $\sup \Delta = \lambda_{\sup} \equiv \sup_{\psi} \lambda.$ As of today, the exact value of $\lambda_{\sup}$ remains unknown, while the most accurate numerical estimate stands at \cite{PGKW06new}
\begin{equation}
	\lambda_{\sup} \simeq 0.0384517 \,.
\label{BM_const}
\end{equation}
It is intriguing that the constant $\lambda_{\sup}$, called the Bracken-Melloy constant, is independent not only of the particle mass and of times $\tau$ and $T$, but also of the Planck constant $\hbar$. This observation prompted Bracken and Melloy to suggest that $\lambda_{\sup}$ is a ``new dimensionless quantum number'' that ``reflects the structure of Schr\"odinger's equation, rather than the values of the parameters appearing there''~\cite{BM94Probability, BM14Waiting}. (It has been later pointed out in Ref.~\cite{YHHW12Analytical} that there is a conceptual analogy between, on the one hand, the $\hbar$-independence of $\lambda_{\sup}$ and, on the other hand, the $\hbar$-independence of the reflection coefficient in the problem of quantum-mechanical scattering off a step potential.)

Following the pioneering work of Bracken and Melloy~\cite{BM94Probability}, numerous studies of the QB effect have been reported in the literature. Accurate numerical estimates of the Bracken-Melloy constant were obtained in Refs.~\cite{EFV05Quantum, PGKW06new}. Analytical examples of states with large QB were constructed in Refs.~\cite{YHHW12Analytical, HGL+13Quantum}. The relation between QB and the arrival-time problem was discussed in Refs.~\cite{MPL99Arrival, ML00Arrival, HBLO19Quasiprobability}. Some aspects of the spatial extent of QB were studied in Refs.~\cite{EFV05Quantum, Ber10Quantum, BCL17Quantum}. Probability backflow in quantum systems with rotational motion, such as an electron in a constant magnetic field, was addressed in Ref.~\cite{Str12Large}. A scheme for observing QB in experiments with Bose-Einstein condensates was proposed in Ref.~\cite{PTMM13Detecting}. QB has been also investigated under the action of a constant force~\cite{MB98velocity}, in the presence of spin-orbit coupling~\cite{MPM+14Interference}, thermal noise~\cite{AGP16Quantum}, dissipations~\cite{MM20Dissipative}, and in scattering situations~\cite{BCL17Quantum}. QB in relativistic quantum mechanics was studied in Refs.~\cite{MB98Probability, SC18Quantum, ALS19Relativistic}. While an experimental observation of QB in a truly {\it quantum} system is still missing, an {\it optical} equivalent of the effect has been recently realized in the laboratory~\cite{EZB20Observation}.

Recently, it has been shown \cite{Gou19Equivalence} that classically-forbidden probability transfer may also occur in a seemingly different problem of quantum reentry (QR) that can be formulated as follows. Suppose that initially, at $t = 0$, the particle is localized on the negative position semi-axis, $x < 0$, so that $P_<(0) = 1$; the particle momentum $p$ is unconstrained. At later times, $t > 0$, as a result of free motion the particle may cross the origin and enter the region $x > 0$, yielding $P_<(t) < 1$ for $t > 0$. Just as in the case of QB, the quantity of interest is the probability transfer $\Delta(\tau, T)$, defined by Eq.~\eqref{Delta_def}, with the only modification that one now requires not only $T > 0$, but also $\tau > 0$. In classical mechanics, once the free particle has left the $x < 0$ region, it can no longer reenter the region, meaning that the classical-mechanical value of $\Delta(\tau, T)$ cannot be positive -- a classical reentry is impossible. The situation is different in quantum mechanics: there exist states, initially localized in the $x < 0$ region, for which $\Delta(\tau, T)$ is positive. This is the manifestation of the QR effect. QR may take place not only in free space, as described here, but also in the presence of an external potential, e.g., when a particle ``leaks'' out of a quasi-stable trap through a $\delta$-potential barrier~\cite{DT19Decay}.

Interestingly, the least upper bound on the QR probability appears to equal the Bracken-Melloy constant, $\lambda_{\sup}$ \cite{Gou19Equivalence}. This suggests the existence of a deep connection between QB and QR. In this paper, we elucidate this connection by showing that both QB and QR effects can be viewed as special cases of a generalized backflow problem. More surprisingly, we show that the least upper bound on the classically-forbidden probability transfer in this generalized backflow problem exceeds the Bracken-Melloy constant.

This paper is organized as follows. In Section~\ref{Sec:QB-QR} we discuss the phase-space interpretations of QB and QR, and make a connection between the two effects. In Section~\ref{Sec:general} we formulate a generalized backflow problem, and study in detail one particular example. We summarize and discuss our findings in Section~\ref{Sec:end}. Some calculations are deferred to the Appendixes.

\section{A unified view on quantum backflow and quantum reentry}
\label{Sec:QB-QR}

In what follows, we denote the position and momentum operators by $\hat{x}$ and $\hat{p}$, and their corresponding eigenstates by $| x \rangle$ and $| p \rangle$, respectively. The eigenstates are normalized as $\langle x | x' \rangle = \delta(x - x')$ and $\langle p | p' \rangle = \delta(p - p)$, where $\delta(\cdot)$ is the Dirac $\delta$-function. We consider the motion of a quantum particle of mass $m$ under the action of the free-space Hamiltonian
\begin{equation}
	\hat{H} = \frac{\hat{p}^2}{2 m} \,.
\label{Hamiltonian}
\end{equation}
The corresponding evolution operator is given by
\begin{equation}
	\hat{U}(t) = \exp \left(-\frac{i t}{2 \hbar m} \hat{p}^2 \right) \,.
\label{U_def}
\end{equation}

\subsection{Qualitative relation between quantum backflow and quantum reetry}

\begin{figure}[h]
	\centering
	\includegraphics[width=0.35\textwidth]{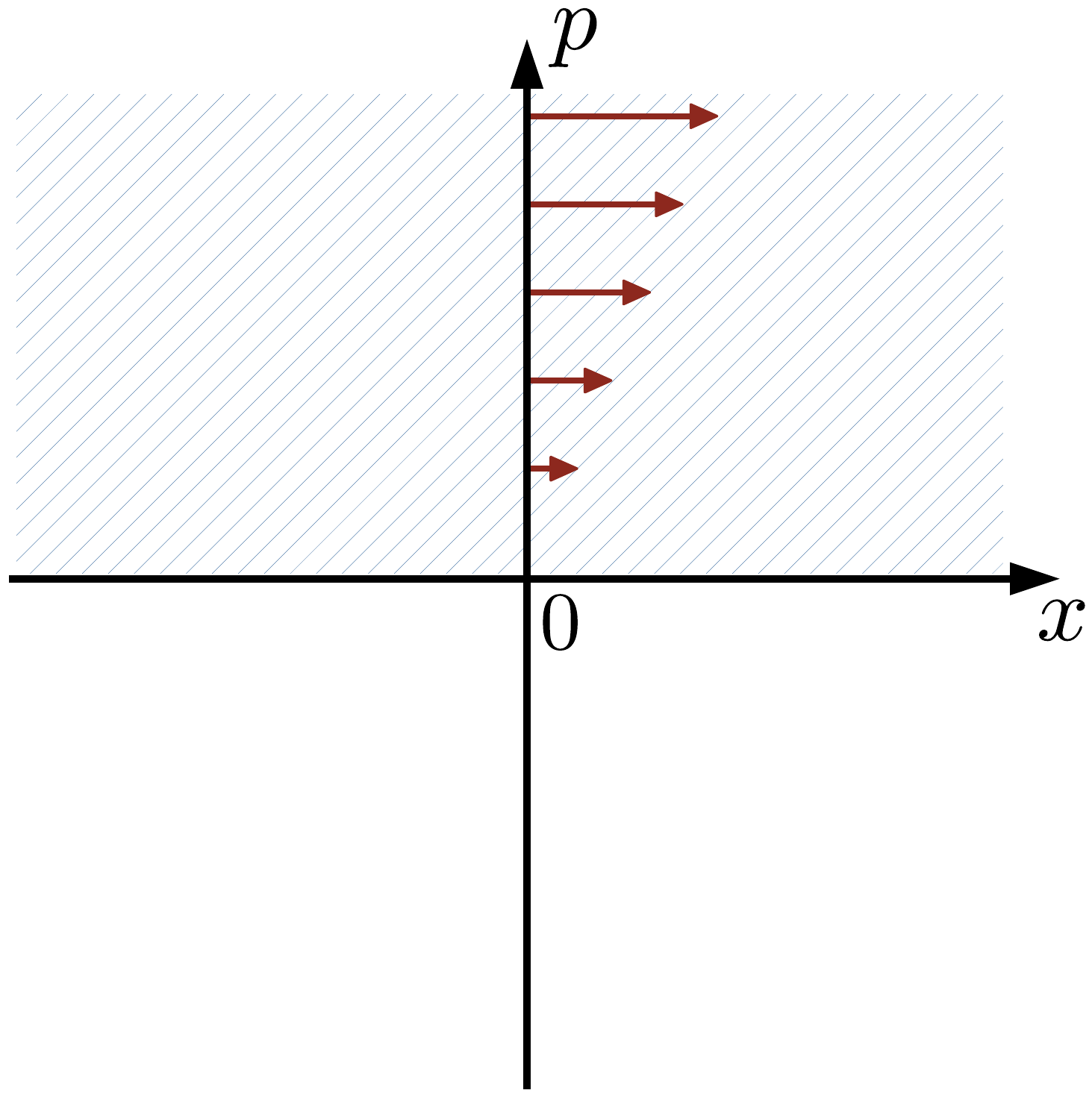}
	\caption{Classical-mechanical representation of a particle with a positive momentum, $p > 0$. The phase-space probability density is supported by the upper half-plane (hatched area). The probability flow through the spatial point $x = 0$ is parallel to the position axis (red arrows).}
	\label{fig1}
\end{figure}
In the QB setting, the state of a particle at time $t = 0$ has the form
\begin{equation}
	\int_0^{\infty} dp \, f(p) | p \rangle \,,
\end{equation}
where $f$ is a complex-valued function normalized as $\int_0^{\infty} dp \, |f(p)|^2 = 1$. The state of the particle during the time interval $\tau < t < \tau + T$, relevant to backflow analysis (see Eq.~\eqref{Delta_def}), is given by
\begin{equation}
	\int_0^{\infty} dp \, f(p) \hat{U}(t) | p \rangle = \int_0^{\infty} dp \, f(p) e^{-i p^2 t / 2 \hbar m} | p \rangle \,.
\end{equation}
It is clear that any momentum measurement performed on this state is bound to give a positive result, $p > 0$, with the probability density $|f(p)|^2$. So, from the classical-mechanical viewpoint, the particle is located in the upper half-plane of phase space, and any probability transfer through the spatial point $x = 0$ may only occur in the left-to-right direction (see Fig.~\ref{fig1}). From the quantum-mechanical perspective however the fact that the outcome of a measurement of $\hat{p}$ is (at all times) certain to be positive does not prevent the right-to-left probability transfer $\Delta(\tau,T)$ from being positive.

\begin{figure}[h]
	\centering
	\includegraphics[width=0.35\textwidth]{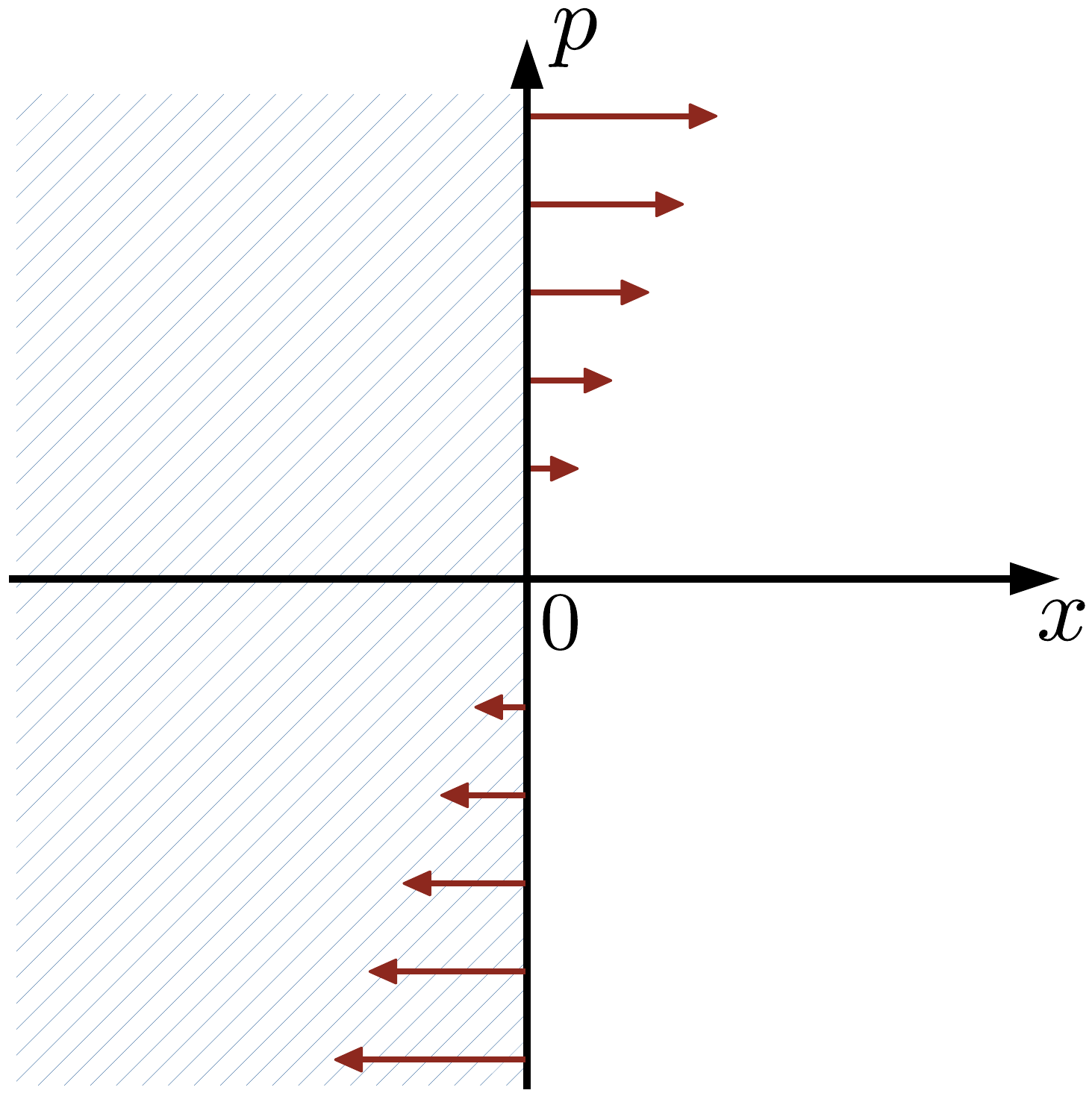}
	\caption{Classical-mechanical representation of a particle with a negative position, $x < 0$. The phase-space probability density is supported by the left half-plane (hatched area). The instantaneous probability flow through the spatial point $x = 0$ is illustrated with red arrows.}
	\label{fig2}
\end{figure}
\begin{figure}[h]
	\centering
	\includegraphics[width=0.39\textwidth]{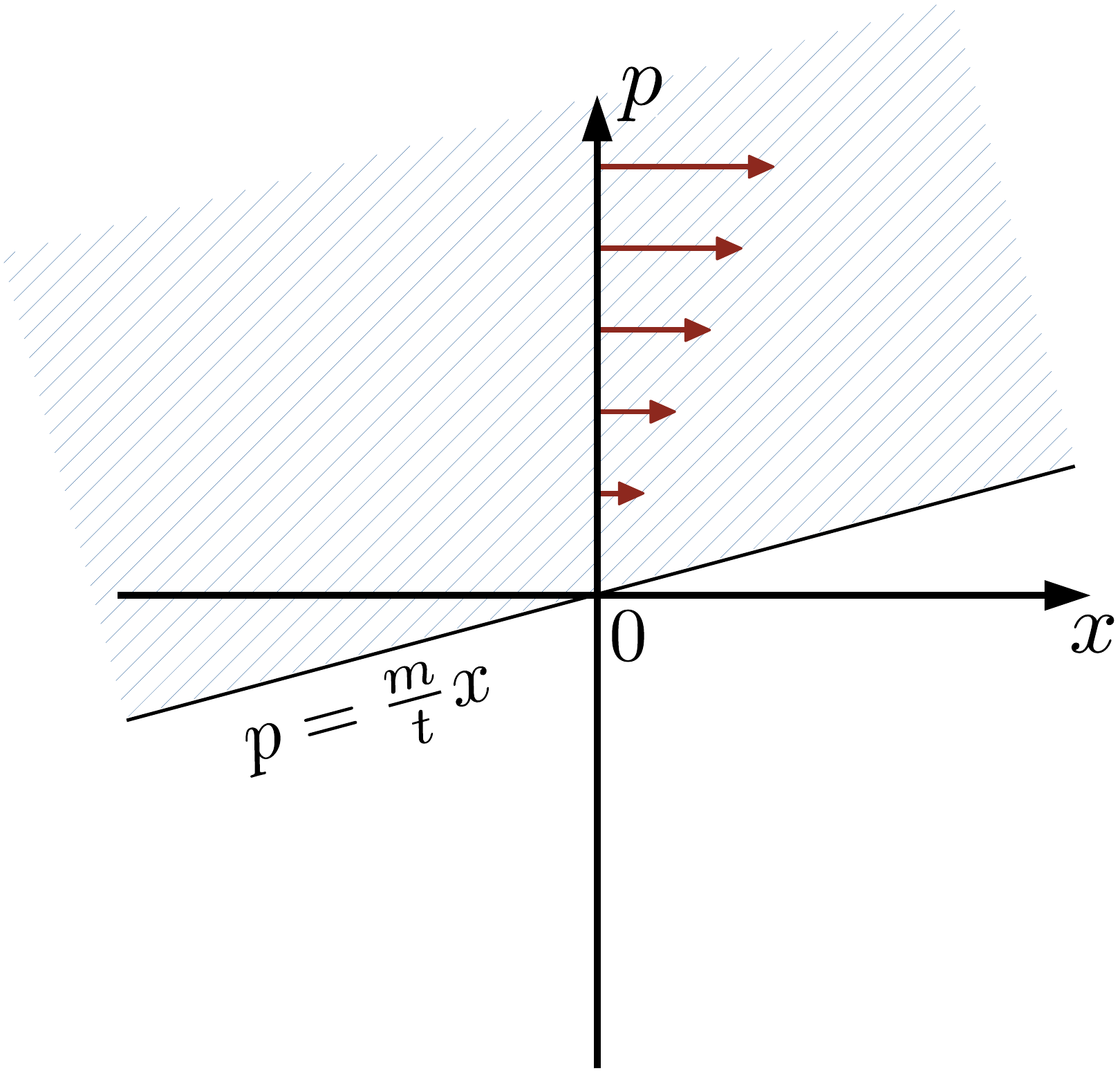}
	\caption{Classical-mechanical representation at $t > 0$ of a particle that was initially (at $t = 0$) confined to the region of negative positions. The phase-space probability density is supported by the half-plane above the line $p = \frac{m}{t} x$ (hatched area). The probability flow through the spatial point $x = 0$ is parallel to the position axis (red arrows).}
	\label{fig3}
\end{figure}
Let us now look at the QR problem. At time $t = 0$, the particle is localized in the region of negative positions (see Fig.~\ref{fig2}). This means that the corresponding quantum state has the form
\begin{equation}
	\int_{-\infty}^0 dx \, g(x) | x \rangle \,,
\end{equation}
where $g$ is some complex-valued function normalized as $\int_{-\infty}^0 dx \, |g(x)|^2 = 1$. A measurement of $\hat{x}$ performed on this state is guaranteed to return a negative result, $x < 0$, with the probability density $|g(x)|^2$. In the course of its evolution through time $t > 0$, the particle state becomes
\begin{equation}
	\int_{-\infty}^0 dx \, g(x) \hat{U}(t) | x \rangle \,.
\label{QR_t>0_state}
\end{equation}
It is straightforward to show (see Appendix~\ref{App:U|x>}) that $\hat{U}(t) | x \rangle$ is an eigenstate of the Hermitian operator $\hat{p} - \frac{m}{t} \hat{x}$, with the corresponding eigenvalue equal to $-\frac{m x}{t}$, i.e.
\begin{equation}
	\left( \hat{p} - \frac{m}{t} \hat{x} \right) \hat{U}(t) | x \rangle = -\frac{m x}{t} \hat{U}(t) | x \rangle \,.
\label{U|x>}
\end{equation}
Therefore, a measurement of $\hat{p} - \frac{m}{t} \hat{x}$, performed on the state given by Eq.~\eqref{QR_t>0_state}, is guaranteed to give a positive result. From the classical-mechanical viewpoint, this means that the particle at time $t >0$ is located above the $p = \frac{m}{t} x$ line in phase space (see Fig~\ref{fig3}). Therefore, according to the laws of classical mechanics, any probability flow through the spatial point $x = 0$ may only take place in the left-to-right direction. However, the quantum-mechanical analysis of this problem presented below shows that the system {\it can} give rise to a positive (classically-forbidden) probability transfer in the right-to-left direction, i.e. one might have $\Delta(\tau, T) > 0$ with $\tau, T > 0$.

Figures~\ref{fig1} and \ref{fig3} elucidate the connection between QB and QR: both phenomena can be regarded as classically-forbidden probability flow for quantum states with position-momentum correlations. In general, one is interested in the right-to-left probability transfer, $\Delta(\tau, T)$, produced by a quantum state whose classical phase-space probability density, at $t = \tau$, vanishes below the line $p = \frac{m}{\tau} x$. The QR problem corresponds to the case of a finite $\tau > 0$. The QB problem is recovered in the limit $\tau \to \infty$.

\subsection{Probability transfer operator}

We begin our analysis of backflow for position-momentum correlated states by introducing the probability transfer operator $\hat{D}$. The right-to-left probability transfer $\Delta(\tau, T)$, defined by Eq.~\eqref{Delta_def}, can be written as the following expectation value:
\begin{equation}
	\Delta(\tau, T) = \langle \tau | \hat{D}(T) | \tau \rangle \,,
\label{Delta_in_terms_of_D}
\end{equation}
where $| \tau \rangle$ is the particle state at time $t = \tau$, and
\begin{equation}
	\hat{D}(t) = \hat{U}^{\dag}(t) \Theta(-\hat{x}) \hat{U}(t) - \Theta(-\hat{x}) \,.
\label{D_1}
\end{equation}
Here, $\Theta(\cdot)$ is the Heaviside step function, and the dagger $\dag$ denotes Hermitian conjugation. It is convenient to rewrite the probability transfer operator in a symmetric form as
\begin{equation}
	\hat{D}(t) = \hat{U}^{\dag}( \tfrac{t}{2} ) \hat{B}(t) \hat{U}( \tfrac{t}{2} )
\label{D_2}
\end{equation}
with
\begin{equation}
	\hat{B}(t) = \hat{U}^{\dag}(\tfrac{t}{2}) \Theta(-\hat{x}) \hat{U}(\tfrac{t}{2}) - \hat{U}(\tfrac{t}{2}) \Theta(-\hat{x}) \hat{U}^{\dag}(\tfrac{t}{2}) \,.
\label{B_def}
\end{equation}
It is clear from their definitions that both $\hat{B}$ and $\hat{D}$ are Hermitian operators. A straightforward calculation (see Appendix~\ref{App:B_op}) yields the following momentum representation of $\hat{B}$:
\begin{equation}
	\langle p | \hat{B}(t) | p' \rangle = -\frac{1}{\pi (p - p')} \sin \left[ \frac{t}{4 \hbar m} \left( p^2 - {p'}^2 \right) \right] \,.
\label{B_mom_rep}
\end{equation}
The function in the right-hand side of the last equation is the same as the backflow kernel function originally derived by Bracken and Melloy~\cite{BM94Probability}. Finally, in view of Eqs.~\eqref{D_2} and \eqref{B_mom_rep}, we have
\begin{align}
	\langle p | \hat{D}(t) | p' \rangle = -&\frac{1}{\pi (p - p')} \sin \left[ \frac{t}{4 \hbar m} \left( p^2 - {p'}^2 \right) \right] \nonumber \\
	&\times \exp \left[ \frac{i t}{4 \hbar m} \left( p^2 - {p'}^2 \right) \right] \,.
\label{D_mom_rep}
\end{align}
This representation of $\hat{D}$ will be used below.

\subsection{Supremum of classically-forbidden probability transfer}

We now address the right-to-left probability transfer $\Delta(\tau, T)$, given by Eq.~\eqref{Delta_in_terms_of_D}, for states $| \psi \rangle$ with position-momentum correlations of the type illustrated in Fig.~\ref{fig3}. More specifically, we consider the particle state at time $t = \tau$ to have the form
\begin{equation}
	| \tau \rangle = \int_0^{\infty} dz \, \Psi(z) | z \rangle \,,
\label{psi_z_rep}
\end{equation}
where $\Psi$ is a complex-valued function of a real variable, and states $| z \rangle$ are orthonormal eigenstates of the Hermitian operator $\hat{p} - k \hat{x}$ with $k = \frac{m}{\tau} > 0$, i.e.
\begin{equation}
	\left( \hat{p} - k \hat{x} \right) | z \rangle = z | z \rangle \,,
\label{z_def}
\end{equation}
with
\begin{equation}
	\langle z | z' \rangle = \delta(z - z') \,.
\label{z_norm}
\end{equation}
Since, in momentum representation, $\hat{x}$ is given by $i \hbar \frac{\partial}{\partial p}$, it is easy to see that
\begin{equation}
	\langle p | z \rangle = \frac{1}{\sqrt{2 \pi \hbar k}} \exp \left( -\frac{i}{2 \hbar k} p^2 + \frac{i z}{\hbar k} p \right) \,.
\label{z_mom_rep}
\end{equation}
The normalization condition $\langle \tau | \tau \rangle = 1$ imposes the following constraint on $\Psi$:
\begin{equation}
	\int_0^{\infty} dz \, |\Psi(z)|^2 = 1 \,.
\label{Psi_norm}
\end{equation}

The right-to-left probability transfer, generated by $| \tau \rangle$ during the time interval $\tau < t < \tau + T$, is obtained by substituting Eq.~\eqref{psi_z_rep} into Eq.~\eqref{Delta_in_terms_of_D}:
\begin{equation}
	\Delta(\tau,T) = \int_0^{\infty} dz \int_0^{\infty} dz' \, \Psi^*(z) \langle z | \hat{D}(T) | z' \rangle \Psi(z') \,.
\label{Delta_vs_<z|D|z'>}
\end{equation}
A straightforward calculation yields (see Appendix~\ref{App:<z|D|z'>}, or, alternatively, follow the method adopted in Section~\ref{Sec:general})
\begin{equation}
	\langle z | \hat{D}(T) | z' \rangle = -\frac{1}{\pi} e^{-i \alpha z^2} \frac{\sin \left[ \beta \left( z^2 - {z'}^2 \right) \right]}{z - z'} e^{i \alpha {z'}^2} \,,
\label{<z|D|z'>}
\end{equation}
with
\begin{equation}
	\alpha = \frac{1}{4 \hbar k} \left( 1 + \frac{1}{1 + k T / m} \right)
\label{alpha}
\end{equation}
and
\begin{equation}
	\beta = \frac{1}{4 \hbar k} \left( 1 - \frac{1}{1 + k T / m} \right) \,.
\label{beta}
\end{equation}
Then, substituting Eq.~\eqref{<z|D|z'>} into Eq.~\eqref{Delta_vs_<z|D|z'>}, and defining
\begin{equation}
	\psi(u) = \beta^{-1/4} \exp \left( -\frac{i \alpha}{\beta} u^2 \right) \Psi \left( \frac{u}{\sqrt{\beta}} \right) \,,
\end{equation}
we obtain
\begin{equation}
	\Delta(\tau, T) = -\frac{1}{\pi} \int_0^{\infty} du \int_0^{\infty} du' \, \psi^*(u) \frac{\sin \left( u^2 - {u'}^2 \right)}{u - u'} \psi(u') \,.
\label{Delta_in_terms_of_varphi}
\end{equation}
The normalization condition, Eq.~\eqref{Psi_norm}, now reads
\begin{equation}
	\int_0^{\infty} du \, |\psi(u)|^2 = 1 \,.
\label{varphi_norm}
\end{equation}

The supremum of the classically-forbidden right-to-left probability transfer is obtained by optimizing $\Delta(\tau, T)$, Eq.~\eqref{Delta_in_terms_of_varphi}, subject to the normalization constraint on $\psi$, Eq.~\eqref{varphi_norm}. This is the variational problem originally considered in Ref.~\cite{BM94Probability}. The corresponding Euler-Lagrange equation is given by Eq.~\eqref{int_eig_prob}. The supremum of $\Delta(\tau, T)$ equals the Bracken-Melloy constant, Eq.~\eqref{BM_const}, and is independent of the slope $k = \frac{m}{\tau}$ characterizing the position-momentum correlation of the initial state (see Eqs.~\eqref{psi_z_rep} and \eqref{z_def}). In other words, the supremum of the right-to-left probability transfer is the same in the QB and QR problems.

\section{Generalized backflow problem}
\label{Sec:general}

In the previous section, we have shown that both QB and QR effects manifest themselves as (classically forbidden) right-to-left probability transfer for initial states with linear position-momentum correlations, for which the measurement of $\hat{p} - k \hat{x}$, with $k \ge 0$, is guaranteed to yield a positive result. We now extend our study to states with nonlinear position-momentum correlations, for which the outcome of measuring $S(\hat{p}) - \hat{x}$ is certain to be positive. The real function $S$ is such that the curve $S(p) - x = 0$ does not intersect the fourth quadrant of the phase-space plane, i.e. the intersection of $\{ (x,p) : S(p) - x > 0 \}$ and $\{ (x,p) : x > 0 \; \& \; p < 0 \}$ is empty. Clearly, from the classical-mechanical viewpoint, the fact that the support of the phase-space density of a state has no overlap with $\{ (x,p) : x > 0 \; \& \; p < 0 \}$ implies that the state cannot give rise to any positive right-to-left probability transfer.

More concretely, we consider particle states $| \tau \rangle$ of the form
\begin{equation}
	| \tau \rangle = \int_0^{\infty} dw \, \Phi(w) | w \rangle \,,
\label{phi_w_rep}
\end{equation}
where $\Phi$ is a complex-valued function of a real variable, and states $| w \rangle$ are orthonormal eigenstates of the Hermitian operator $S(\hat{p}) - \hat{x}$ \footnote{In general, the sum of two unbounded Hermitian operators is not necessarily Hermitian. However, the operator $S(\hat{p}) - \hat{x}$ is a unitary transformation of the operator $-\hat{x}$, and as such is Hermitian. Indeed, it is straightforward to check that $S(\hat{p}) - \hat{x} = \hat{U}^{-1} (-\hat{x}) \hat{U}$, with $\hat{U} = e^{i R(\hat{p}) / \hbar}$ and $R(p) = \int^{p} dq \, S(q)$ (cf.~Eq.~\eqref{w_mom_rep})}, i.e.
\begin{equation}
	\big( S(\hat{p}) - \hat{x} \big) | w \rangle = w | w \rangle \,,
\label{w_def}
\end{equation}
with
\begin{equation}
	\langle w | w' \rangle = \delta(w - w') \,.
\label{w_norm}
\end{equation}
In momentum representation, $\hat{x}$ is given by $i \hbar \frac{\partial}{\partial p}$, and so
\begin{equation}
	\langle p | w \rangle = \frac{1}{\sqrt{2 \pi \hbar}} \exp \left( -\frac{i}{\hbar} \int^p dq \, S(q) + i \frac{w p}{\hbar} \right) \,.
\label{w_mom_rep}
\end{equation}
The normalization condition $\langle \tau | \tau \rangle = 1$ imposes the following constraint on $\Phi$:
\begin{equation}
	\int_0^{\infty} dw \, |\Phi(w)|^2 = 1 \,.
\label{Phi_norm}
\end{equation}
The right-to-left probability transfer during the time interval $\tau < t < \tau + T$ is given by Eq.~\eqref{Delta_in_terms_of_D}, which, using the $w$-representation, can be written as
\begin{equation}
	\Delta(\tau,T) = \int_0^{\infty} dw \int_0^{\infty} dw' \, \Phi^*(w) \langle w | \hat{D}(T) | w' \rangle \Phi(w') \,.
\label{Delta_vs_<w|D|w'>}
\end{equation}
Repeating the steps that led to Eq.~\eqref{App:<z|D|z'>:eq0}, we obtain the following expression for the kernel $\langle w | \hat{D}(T) | w' \rangle$:
\begin{widetext}
\begin{align}
	\langle w | \hat{D}(T) | w' \rangle
	= -\frac{1}{2 \pi^2 \hbar} \int_{-\infty}^{+\infty} dp \int_{-\infty}^{+\infty} dp' &\, \frac{\sin \left[ \frac{T}{4 \hbar m} \left( p^2 - {p'}^2 \right) \right]}{p - p'} \nonumber \\
	&\times 
	\exp \left[ \frac{i}{\hbar} \int_{p'}^p dq \, S(q) + i \frac{T}{4 \hbar m} \left( p^2 - {p'}^2 \right) - i \frac{w p - w' p'}{\hbar} \right] \,.
\label{<w|D|w'>}
\end{align}

It is convenient to introduce the following dimensionless versions of the functions $\Phi$ and $S$:
\begin{equation}
	 \phi(u) = \left( \frac{\hbar T}{4 m} \right)^{1/4} \Phi \left(u \sqrt{\frac{\hbar T}{4 m}} \right) \,, \qquad s(u) = \sqrt{\frac{4 m}{\hbar T}} \, S \left( u \sqrt{\frac{4 \hbar m}{T}} \right) \,.
\label{dimensionless_variables}
\end{equation}
This corresponds to taking $\sqrt{\frac{\hbar T}{4 m}}$ for the unit of position and $\sqrt{\frac{4 \hbar m}{T}}$ for the unit of momentum. Now the expression for the right-to-left probability transfer, Eqs.~\eqref{Delta_vs_<w|D|w'>} and \eqref{<w|D|w'>}, takes the form
\begin{equation}
	\Delta(\tau, T) = \int_0^{\infty} du \int_0^{\infty} du' \, \phi^*(u) K(u,u') \phi(u') \,,
\label{Delta_dimless}
\end{equation}
with
\begin{equation}
	K(u, u') = -\frac{1}{2 \pi^2} \int_{-\infty}^{\infty} d\xi \int_{-\infty}^{\infty} d\xi' \, \frac{\sin \left( \xi^2 - {\xi'}^2 \right)}{\xi - \xi'} \exp \left[ i \int_{\xi'}^{\xi} dp \, s(p) + i \left( {\xi}^2 - {\xi'}^2 \right) - i u \xi + i u' \xi' \right] \,.
\label{K_def}
\end{equation}
\end{widetext}
We also note that the normalization condition for $\Phi$, given by Eq.~\eqref{Phi_norm}, translates into
\begin{equation}
	\int_0^{\infty} du \, |\phi(u)|^2 = 1 \,.
\label{phi_norm}
\end{equation} 

Optimization of $\Delta(\tau, T)$ with respect to $\phi$, subject to the normalization constraint, Eq.~\eqref{phi_norm}, yields the following eigenproblem:
\begin{equation}
	\int_0^{\infty} du' \, K(u, u') \phi(u') = \mu \phi(u) \,.
\label{K_eigprob}
\end{equation}
The supremum of $\Delta$ is given by the supremum of the eigenvalue spectrum, $\mu_{\sup} = \sup_{\phi} \mu$. It is easy to show that in the case of a straight boundary, $s(p) = p/k$ with $k > 0$, this eigenproblem reduces to the Bracken-Melloy one, Eq.~\eqref{int_eig_prob}, and in this case $\mu_{\sup}$ coincides with $\lambda_{\sup}$. However, for curved boundaries $s(p) - x = 0$, the supremum of the right-to-left probability transfer, $\mu_{\sup}$, generally differs from the Bracken-Melloy constant, $\lambda_{\sup}$. Moreover, as we argue below, one can choose $s(p)$ such that $\mu_{\sup} > \lambda_{\sup}$.

\subsection{Small deformation of a straight phase-space boundary}

Let us consider the case when the phase-space boundary curve $s(p) - x = 0$ is only a small deformation of a straight line:
\begin{equation}
	s(p) = \frac{p}{k} + \epsilon s_1(p) \,,
\end{equation}
where $k > 0$, parameter $\epsilon$ is defined on an interval containing zero, and $s_1(p)$ is a bounded function, such that the curve $s(p) - x = 0$ does not cross the fourth quadrant of the phase space. (A concrete example of such function will be considered in Section~\ref{Sec:Ex}.) Assuming continuity around $\epsilon = 0$, we write
\begin{align*}
	K(u,u') &= K_0(u,u') + \epsilon K_1(u,u') + O(\epsilon^2) \,, \\
	\phi(u) &= \phi_0(u) + \epsilon \phi_1(u) + O(\epsilon^2) \,, \\
	\mu &= \mu_0 + \epsilon \mu_1 + O(\epsilon^2) \,.
\end{align*}
Substituting these expansions into Eq.~\eqref{K_eigprob} and comparing terms of the same order in $\epsilon$, we obtain
\begin{equation}
	\int_0^{\infty} du' \, K_0(u, u') \phi_0(u') = \mu_0 \phi_0(u)
\label{K0_eigprob}
\end{equation}
and
\begin{align}
	\int_0^{\infty} du' &\Big( K_0(u,u') \phi_1(u') + K_1(u,u') \phi_0(u') \Big) \nonumber \\
	&= \mu_0 \phi_1(u) + \mu_1 \phi_0(u) \,.
\label{K1_eigprob}
\end{align}
Then, we multiply both sides of Eq.~\eqref{K1_eigprob} by $\phi^*_0(u)$, integrate over $u$, and make use of the facts that $K_0(u',u) = \big[ K_0(u,u') \big]^*$ and that $\phi_0$ fulfils Eqs.~\eqref{K0_eigprob} and \eqref{phi_norm}. This yields
\begin{equation}
	\mu_1 = \int_0^{\infty} du \int_0^{\infty} du' \, \phi_0^*(u) K_1(u,u') \phi_0(u') \,.
\label{mu_1}
\end{equation}
In fact, Eq.~\eqref{mu_1} is nothing but the prediction of the standard non-degenerate perturbation theory for the eigenvalue spectrum of a linear Hermitian operator.

We now put forward the following argument. Let us suppose that $\phi_0(u)$ is a normalized eigenfunction corresponding the eigenvalue $\mu_0 = \lambda_{\sup}$. Then, for small $\epsilon$, we expect to have
\begin{equation}
	\mu_{\sup} \simeq \lambda_{\sup} + \epsilon \mu_1 \,,
\label{mu_sup_linear_approx}
\end{equation}
where $\mu_1$ is determined by Eq.~\eqref{mu_1}. In general, there is no reason for $\mu_1$ to vanish. In turn, the assumption $\mu_1 \not= 0$ implies that there is a value of $\epsilon$ for which $\mu_{\sup} > \lambda_{\sup}$. Below we demonstrate that this intuitive argument is indeed valid by numerically computing $\mu_{\sup}$ for a specific example of a phase-space boundary curve.

\subsection{Example}
\label{Sec:Ex}

\begin{figure}[h]
	\centering
	\includegraphics[width=0.5\textwidth]{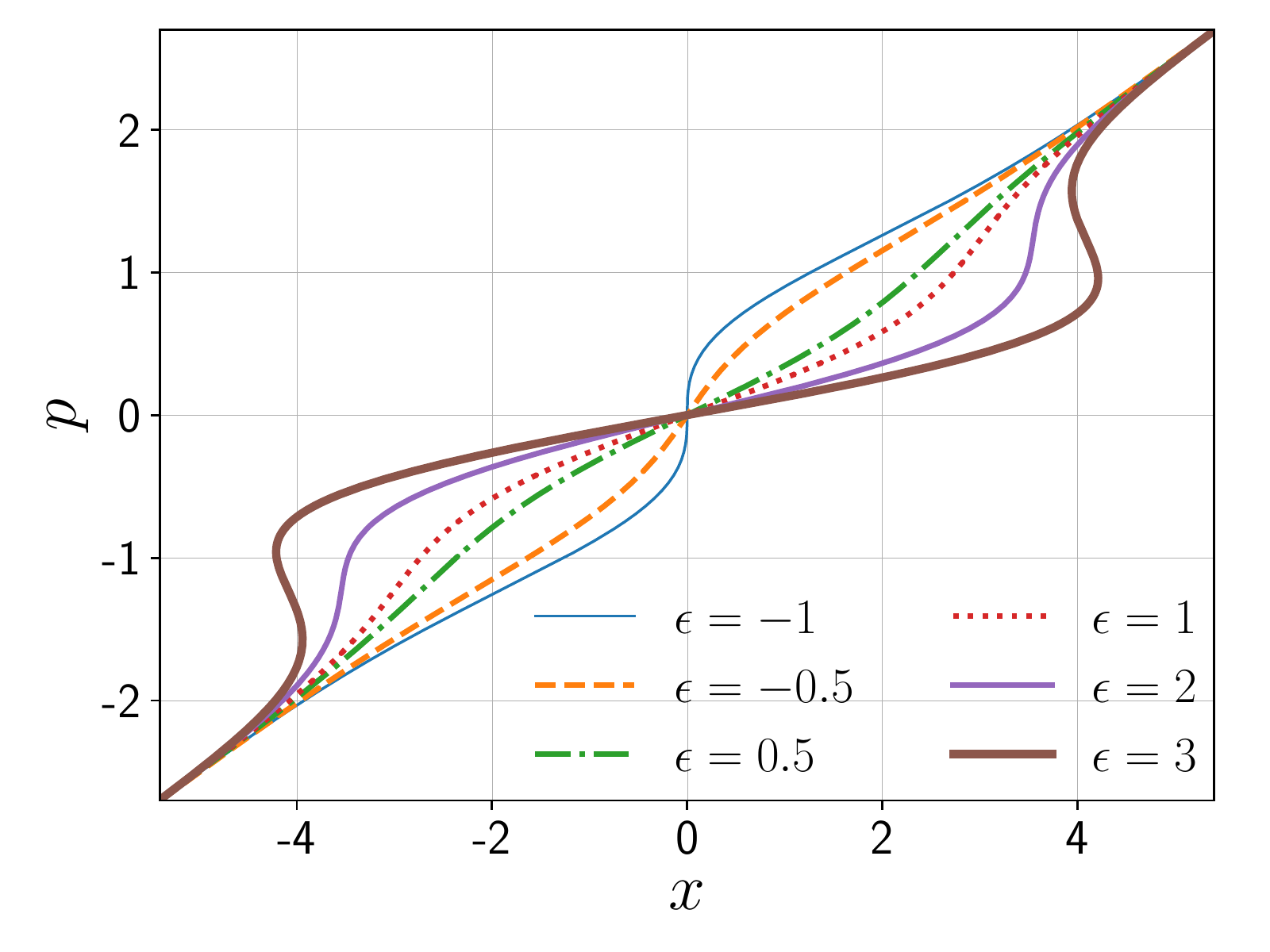}
	\caption{Phase-space boundary curve $s(p) - x = 0$, with $s(p)$ given by Eq.~\eqref{s}, for $\epsilon = -1, \, -0.5, \, 0.5, \, 1, \, 2, \, 3$. Here, both $x$ and $p$ are dimensionless variables (see Eq.~\eqref{dimensionless_variables} for the rescaling).}
	\label{fig4}
\end{figure}

Let us consider the case
\begin{equation}
	s(p) = 2 p \left( 1 + \epsilon e^{-p^2} \right) \,.
\label{s}
\end{equation}
Provided $\epsilon > -1$, the curve $s(p) - x = 0$ does not penetrate the fourth quadrant of the phase-space plane (see Fig.~\ref{fig4}). Our aim is to demonstrate that, for some values of $\epsilon$, the supremum of the right-to-left probability transfer, $\mu_{\sup}$, exceeds the Bracken-Melloy constant, $\lambda_{\sup}$.

We start by recasting the kernel $K(u,u')$, defined by Eq.~\eqref{K_def}, in a form more convenient for the present computation. Substituting the identity
\begin{equation}
	\frac{\sin \left( \xi^2 - {\xi'}^2 \right)}{\xi - \xi'} = \frac{\xi + \xi'}{2} \int_{-1}^{1} d\nu \, \exp \left[ i \left( \xi^2 - {\xi'}^2 \right) \nu \right]
\end{equation}
into Eq.~\eqref{K_def}, and then changing the integration order, we get
\begin{equation}
	K(u, u') = -\frac{i}{4 \pi^2} \int_{-1}^1 d\nu \, \left( \frac{\partial}{\partial u} - \frac{\partial}{\partial u'} \right) I(u, \nu) I^*(u', \nu) \,,
\label{K_new_representation}
\end{equation}
where
\begin{equation}
	I(u,\nu) = \int_{-\infty}^{+\infty} d\xi \, \exp \left( i \int^{\xi} dp \, s(p) + i (1 + \nu) \xi^2 - i u \xi \right) \,.
\end{equation}
In the case of $s(p)$ given by Eq.~\eqref{s}, we have $\int^{\xi} dp \, s(p) = \xi^2 - \epsilon e^{-\xi^2}$, and so
\begin{align}
	I(u, \nu)
	&= \int_{-\infty}^{+\infty} d\xi \, \exp \left( -i \epsilon e^{-\xi^2} + i (2 + \nu) \xi^2 - i u \xi \right) \nonumber \\
	&= \sum_{n = 0}^{\infty} \frac{(-i \epsilon)^n}{n!} \int_{-\infty}^{+\infty} d\xi \, e^{-[n - i (2 + \nu)] \xi^2 - i u \xi} \nonumber \\
	&= \sum_{n = 0}^{\infty} \frac{(-i \epsilon)^n}{n!} \sqrt{\frac{\pi}{n - i (2 + \nu)}} e^{ -\frac{u^2}{4 [n - i (2 + \nu)]} } \,. \label{I_as_sum}
\end{align}
Then, substituting Eq.~\eqref{I_as_sum} into Eq.~\eqref{K_new_representation}, and performing some straightforward manipulations (see Appendix~\ref{App:K_expansion}), we arrive at the following expansion:
\begin{equation}
	K(u, u') = \sum_{n=0}^{\infty} \epsilon^n K_n(u, u') \,,
\label{K_expansion}
\end{equation}
where
\begin{widetext}
\begin{equation}
	K_n(u, u') = -\frac{(-i)^n}{8 \pi} \sum_{k=0}^n \frac{(-1)^k}{(n-k)! k!} \int_1^3 dz \, \left( \frac{u}{z + i (n-k)} + \frac{u'}{z - i k} \right) \frac{\exp \left( -i \frac{u^2}{4 [z + i (n-k)]} + i \frac{{u'}^2}{4 (z - i k)} \right)}{\sqrt{[z + i (n-k)] (z - i k)}} \,.
\label{K_n}
\end{equation}
\end{widetext}
For $n = 0$ (and, consequently, $k = 0$), the integral in Eq.~\eqref{K_n} can be done analytically (see Appendix~\ref{App:K_expansion}). This yields the expression
\begin{equation}
	K_0(u,u') = -\frac{1}{\pi} e^{-i u^2 / 6} \frac{\sin \left[ \frac{1}{12} \left( u^2 - {u'}^2 \right) \right]}{u - u'} e^{i {u'}^2 / 6} \,,
\label{K_0}
\end{equation}
which, as expected, is consistent with Eq.~\eqref{<z|D|z'>}. For $n \ge 1$, the integral and sum in Eq.~\eqref{K_n} have to be evaluated numerically.

The numerical evaluation of $\mu_{\sup}$ is based on the method originally used in Ref.~\cite{BM94Probability}. We consider the following discretized version of Eq.~\eqref{K_eigprob}: $\frac{L}{N} \sum_{l = 0}^N K(u_k, u_l) \phi_{L,N}(u_l) = \mu_{L,N} \phi_{L,N}(u_k)$ with $u_k = \frac{L}{N} k$. The kernel $K(u_k, u_l)$ is computed using Eqs.~\eqref{K_expansion} and \eqref{K_n}. We limit our consideration to $-0.5 < \epsilon < 0.5$. For this range, it appears to be sufficient to retain only terms up to (and including) order $\epsilon^4$ in Eq.~\eqref{K_expansion}. Fixing the value of $L$, we solve the discretized eigenproblem numerically for different values of $N$, and observe the following scaling: $\max \mu_{L,N} \simeq \mu_L + C_L N^{-1}$, where $C_L$ is a constant independent of $N$. Extrapolating to $N \to \infty$, we determine $\mu_L$. Repeating this computation for different values of $L$, we observed that $\mu_L \simeq \mu_{\sup} + C L^{-1}$, where $C$ is a constant independent of $L$. Finally, extrapolating to $L \to \infty$, we find $\mu_{\sup}$.

\begin{figure}[h]
	\centering
	\includegraphics[width=0.5\textwidth]{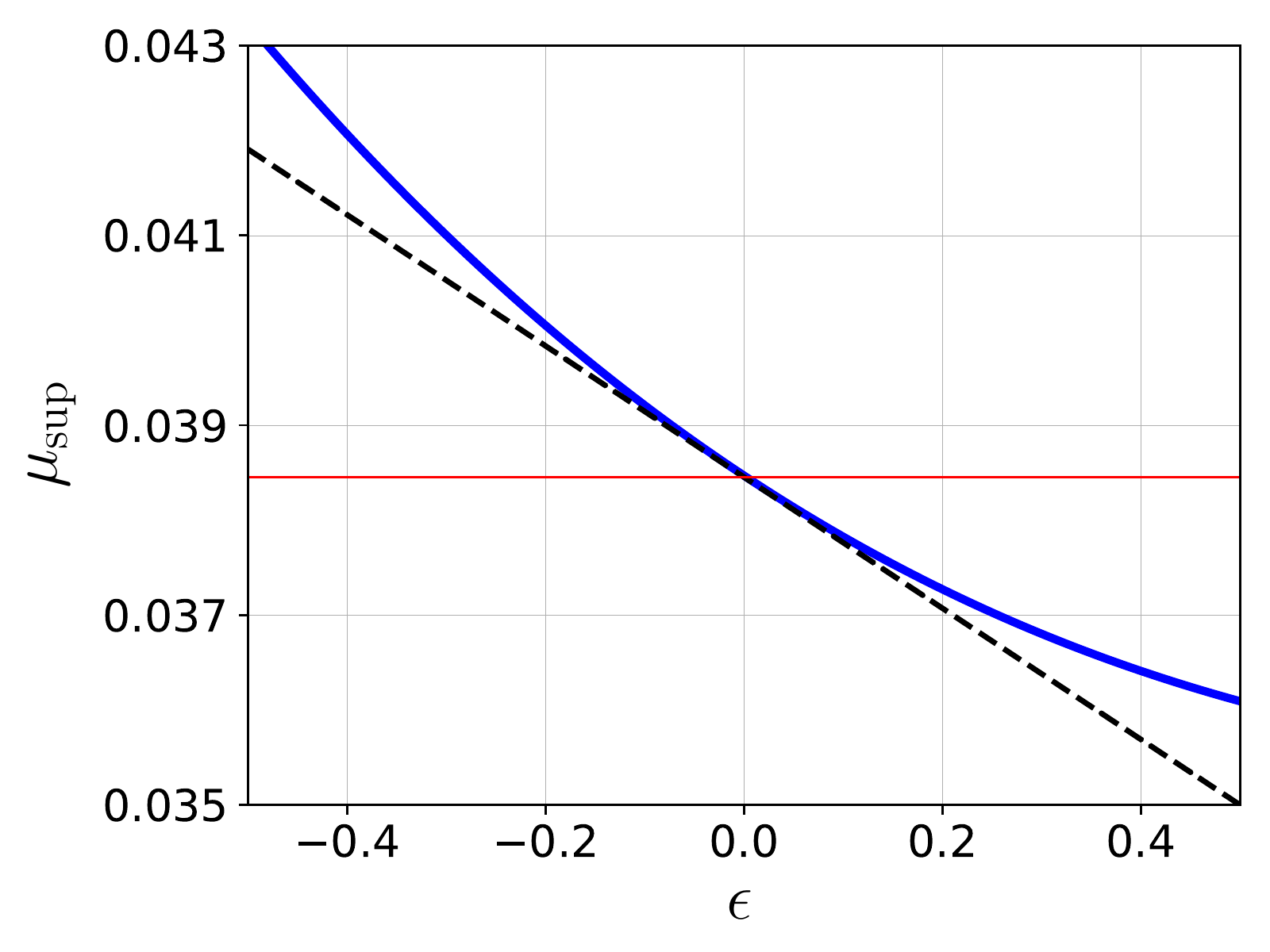}
	\caption{Thick blue curve: The supremum of the classically-forbidden probability transfer, $\mu_{\sup}$, obtained by numerically solving Eq.~\eqref{K_eigprob} for different values of $\epsilon$. (See the text for details.) Thin red line: The Bracken-Melloy constant, $\lambda_{\sup}$. Dashed black line: The linear approximation $\mu_{\sup} = \lambda_{\sup} + \epsilon \mu_1$, with $\mu_1$ computed from Eq.~\eqref{mu_1}.}
	\label{fig5}
\end{figure}

The thick blue curve in Fig.~\ref{fig5} shows the numerically computed function $\mu_{\sup} = \mu_{\sup}(\epsilon)$. The Bracken-Melloy bound, $\lambda_{\sup}$, is presented with the thin red line. The dashed black line shows the linear approximation to $\mu_{\sup}(\epsilon)$, as given by Eq.~\eqref{mu_sup_linear_approx} with the slope evaluated numerically using Eq.~\eqref{mu_1}. The central message conveyed by Fig.~\ref{fig5} is that the classically-forbidden probability transfer from wave packets with nonlinear position-momentum correlations can indeed exceed the Bracken-Melloy bound.

\section{Conclusion}

\label{Sec:end}

The results of this work are twofold. First, we elucidate the reason why the suprema of the classically-impossible probability transfer in the QB and QR problems coincide. We achieve this by showing that both effects can be viewed as a backflow for quantum states with linear position-momentum correlations, defined by Eqs.~\eqref{psi_z_rep} and \eqref{z_def}. For such states, the supremum of the backflow is given by the Bracken-Melloy constant. Second, we formulate a generalized backflow problem for quantum states with nonlineal position-momentum correlations, defined by Eqs.~\eqref{phi_w_rep} and \eqref{w_def}. QB and QR can be viewed as special cases of the generalized backflow. We further present analytical and numerical arguments demonstrating that the supremum of the classically-forbidden probability transfer in the generalized backflow problem exceeds the Bracken-Melloy constant.  

As of today, probability backflow in quantum systems has not been observed experimentally. One of the reasons for this is that, when considered in the original QB formulation, the effect is weak: the QB probability transfer is bounded by less than $3.9\%$ (see Eq.~\eqref{BM_const}). As shown in this paper, the class of quantum states exhibiting classically-forbidden probability flow is larger than that of QB states, and the backflow probability transfer can exceed $3.9\%$. For instance, in the system considered in Section~\ref{Sec:Ex}, more than $4.3\%$ of the total probability can be transported in the ``wrong'' direction (see Fig.~\ref{fig5}). This suggests it could be possible to devise quantum states with large  probability backflow that would facilitate future experiments and offer technological applications.


\appendix

\onecolumngrid

\section{Derivation of Eq.~\eqref{U|x>}}
\label{App:U|x>}

Using the Baker–Campbell–Hausdorff formula,
\begin{equation*}
	e^{\hat{A}} \hat{B} e^{-\hat{A}} = \hat{B} + [\hat{A}, \hat{B}] + \frac{1}{2!} [\hat{A}, [\hat{A}, \hat{B}]] + \frac{1}{3!} [\hat{A}, [\hat{A}, [\hat{A}, \hat{B}]]] + \ldots \,,
\end{equation*}
where $\hat{A}$ and $\hat{B}$ are operators and $[\cdot,\cdot]$ denotes the commutator, we have
\begin{equation*}
	\hat{U}^{-1}(t) \hat{x} \hat{U}(t) = \exp \left(\frac{i t}{2 \hbar m} \hat{p}^2 \right) \hat{x} \exp \left(-\frac{i t}{2 \hbar m} \hat{p}^2 \right) = \hat{x} + \frac{i t}{2 \hbar m} [\hat{p}^2 , \hat{x}] = \hat{x} + \frac{t}{m} \hat{p} \,.
\end{equation*}
From here, it follows that
\begin{equation*}
	\hat{x} \hat{U}(t) = \hat{U}(t) \hat{x} + \frac{t}{m} \hat{p} \hat{U}(t) \,,
\end{equation*}
or
\begin{equation*}
	\left( \hat{p} - \frac{m}{t} \hat{x} \right) \hat{U}(t) = -\frac{m}{t} \hat{U}(t) \hat{x} \,.
\end{equation*}
Applying this operator identity to a position eigenstate $| x \rangle$, we obtain Eq.~\eqref{U|x>}.

\section{Derivation of Eq.~\eqref{B_mom_rep}}
\label{App:B_op}

Starting from Eq.~\eqref{B_def}, and using Eq.~\eqref{U_def}, we write
\begin{align}
	\langle p | \hat{B}(t) | p' \rangle
	&= \langle p | \hat{U}^{\dag}(\tfrac{t}{2}) \Theta(-\hat{x}) \hat{U}(\tfrac{t}{2}) | p' \rangle - \langle p | \hat{U}(\tfrac{t}{2}) \Theta(-\hat{x}) \hat{U}^{\dag}(\tfrac{t}{2}) | p' \rangle \nonumber \\
	&= \left\{ \exp \left[ \frac{i t}{4 \hbar m} \left( p^2 - {p'}^2 \right) \right] - \exp \left[ -\frac{i t}{4 \hbar m} \left( p^2 - {p'}^2 \right) \right] \right\} \langle p | \Theta(-\hat{x}) | p' \rangle \nonumber \\
	&= 2 i \sin \left[ \frac{t}{4 \hbar m} \left( p^2 - {p'}^2 \right) \right] \langle p | \Theta(-\hat{x}) | p' \rangle \,.
\label{App:B_op:eq1}
\end{align}
Then,
\begin{align}
	\langle p | \Theta(-\hat{x}) | p' \rangle
	&= \int_{-\infty}^0 dx \, \langle p | x \rangle \langle x | p' \rangle = \frac{1}{2 \pi \hbar} \int_{-\infty}^0 dx \, e^{-i (p - p') x / \hbar} = \frac{1}{2 \pi} \int_0^{\infty} d\xi \, e^{i (p - p') \xi} \nonumber \\
	&= \frac{1}{2 \pi} \left( \pi \delta(p - p') + i \pv \frac{1}{p - p'} \right) \,,
\label{App:B_op:eq2}
\end{align}
where $\pv$ denotes the Cauchy principal value. Substituting Eq.~\eqref{App:B_op:eq2} into Eq.~\eqref{App:B_op:eq1}, we arrive at Eq.~\eqref{B_mom_rep}.

\section{Derivation of Eq.~\eqref{<z|D|z'>}}
\label{App:<z|D|z'>}

Using Eqs.~\eqref{D_mom_rep} and \eqref{z_mom_rep}, we write
\begin{align}
	\langle z | \hat{D}(T) | z' \rangle
	&= \int_{-\infty}^{+\infty} dp \int_{-\infty}^{+\infty} dp' \, \langle z | p \rangle \langle p | \hat{D}(T) | p' \rangle \langle p' | z' \rangle \nonumber \\
	&= \int_{-\infty}^{+\infty} dp \int_{-\infty}^{+\infty} dp' \, \frac{1}{\sqrt{2 \pi \hbar k}} \exp \left( \frac{i}{2 \hbar k} p^2 - \frac{i z}{\hbar k} p \right) \nonumber \\
	&\hspace*{0.165\textwidth} \times \frac{-1}{\pi (p - p')} \sin \left[ \frac{T}{4 \hbar m} \left( p^2 - {p'}^2 \right) \right] \exp \left[ \frac{i T}{4 \hbar m} \left( p^2 - {p'}^2 \right) \right] \nonumber \\
	&\hspace*{0.165\textwidth} \times \frac{1}{\sqrt{2 \pi \hbar k}} \exp \left( -\frac{i}{2 \hbar k} {p'}^2 + \frac{i z'}{\hbar k} p' \right) \nonumber \\
	&= -\frac{1}{2 \pi^2 \hbar k} \int_{-\infty}^{+\infty} dp \int_{-\infty}^{+\infty} dp' \, \frac{\sin \left[ \frac{T}{4 \hbar m} \left( p^2 - {p'}^2 \right) \right]}{p - p'} \exp \left[ i \frac{1 + k T / 2 m}{2 \hbar k} \left( p^2 - {p'}^2 \right) - i \frac{z p - z' p'}{\hbar k} \right] \,. \label{App:<z|D|z'>:eq0}
\end{align}
Changing the integration variables as
\begin{equation*}
	u = p - p' \,, \qquad v = \frac{p + p'}{2} \,,
\end{equation*}
so that
\begin{equation*}
	p = \frac{u}{2} + v \,, \qquad p' = -\frac{u}{2} + v \,,
\end{equation*}
we obtain
\begin{equation}
	\langle z | \hat{D}(T) | z' \rangle = -\frac{1}{2 \pi^2 \hbar k} \int_{-\infty}^{+\infty} du \, \frac{I(u)}{u} \exp \left( -i \frac{z + z'}{2 \hbar k} u \right) \,,
\label{App:<z|D|z'>:eq1}
\end{equation}
where
\begin{align}
	I(u)
	&= \int_{-\infty}^{+\infty} dv \, \sin \left( \frac{T}{2 \hbar m} u v \right) \exp \left( i \frac{1 + k T / 2 m}{\hbar k} u v - i \frac{z - z'}{\hbar k} v \right) \nonumber \\
	&= \frac{1}{2 i} \left[ \int_{-\infty}^{+\infty} dv \, \exp \left( i \frac{(1 + k T / m) u - z + z'}{\hbar k} v \right) - \int_{-\infty}^{+\infty} dv \, \exp \left( i \frac{u - z + z'}{\hbar k} v \right) \right] \nonumber \\
	&= -i \frac{\pi \hbar k}{1 + k T / m} \delta \left( u - \frac{z - z'}{1 + k T / m} \right) + i \pi \hbar k \delta(u - z + z') \,. \label{App:<z|D|z'>:eq2}
\end{align}
Substituting Eq.~\eqref{App:<z|D|z'>:eq2} into Eq.~\eqref{App:<z|D|z'>:eq1} and evaluating the integral over $u$, we find
\begin{equation*}
	\langle z | \hat{D}(T) | z' \rangle = -\frac{1}{2 \pi i (z - z')} \left[ \exp \left( -i \frac{z^2 - {z'}^2}{2 \hbar k (1 + k T / m)} \right) - \exp \left( -i \frac{z^2 - {z'}^2}{2 \hbar k} \right) \right] \,.
\end{equation*}
Now, introducing constants $\mu$ and $\nu$ such that (cf.~Eqs.~\eqref{alpha} and \eqref{beta})
\begin{equation*}
	\frac{1}{2 \hbar k (1 + k T / m)} = \alpha - \beta \,, \qquad 	\frac{1}{2 \hbar k} = \alpha + \beta \,,
\end{equation*}
we obtain
\begin{equation*}
	\langle z | \hat{D}(T) | z' \rangle = -\frac{e^{-i \alpha \left( z^2 - {z'}^2 \right)}}{\pi (z - z')} \frac{e^{i \beta \left( z^2 - {z'}^2 \right)} - e^{-i \beta \left( z^2 - {z'}^2 \right)}}{2 i} \,,
\end{equation*}
which is equivalent to Eq.~\eqref{<z|D|z'>}.

\section{Derivation of Eq.~\eqref{K_expansion}}
\label{App:K_expansion}

Starting from Eq.~\eqref{I_as_sum}, we have
\begin{align*}
	I(u,\nu) I^*(u',\nu) = &\sum_{n = 0}^{\infty} \frac{(-i \epsilon)^n}{n!} \sqrt{\frac{\pi}{n - i (2 + \nu)}} \exp \left( -\frac{u^2}{4 [n - i (2 + \nu)]} \right) \\
	&\times \sum_{k = 0}^{\infty} \frac{(i \epsilon)^k}{k!} \sqrt{\frac{\pi}{k + i (2 + \nu)}} \exp \left( -\frac{{u'}^2}{4 [k + i (2 + \nu)]} \right) \,.
\end{align*}
Using the Cauchy product formula
\begin{equation*}
	\sum_{n=0}^{\infty} a_n \sum_{k=0}^{\infty} b_k = \sum_{n=0}^{\infty} \sum_{k=0}^n a_{n-k} b_k \,,
\end{equation*}
and the fact that $(-i)^{n-k} i^k = (-1)^{n-k} i^n = (-i)^n (-1)^k$, we obtain
\begin{equation*}
	I(u, \nu) I^*(u', \nu) = \pi \sum_{n=0}^{\infty} (-i \epsilon)^n \sum_{k=0}^n \frac{(-1)^k}{(n-k)! k!} \frac{\exp \left( -\frac{u^2}{4 [n - k - i (2 + \nu)]} - \frac{{u'}^2}{4 [k + i (2 + \nu)]} \right)}{\sqrt{[n - k - i (2 + \nu)] [k + i (2 + \nu)]}}
\end{equation*}
Substituting the last expression into Eq.~\eqref{K_new_representation}, we get
\begin{align}
	K(u, u')
	&= -\frac{i}{4 \pi} \sum_{n=0}^{\infty} (-i \epsilon)^n \sum_{k=0}^n \frac{(-1)^k}{(n-k)! k!} \int_{-1}^1 d\nu \, \left( \frac{\partial}{\partial u} - \frac{\partial}{\partial u'} \right) \frac{\exp \left( -\frac{u^2}{4 [n - k - i (2 + \nu)]} - \frac{{u'}^2}{4 [k + i (2 + \nu)]} \right)}{\sqrt{[n - k - i (2 + \nu)] [k + i (2 + \nu)]}} \nonumber \\
	&= -\frac{1}{8 \pi} \sum_{n=0}^{\infty} (-i \epsilon)^n \sum_{k=0}^n \frac{(-1)^k}{(n-k)! k!} Q_{n,k}(u,u') \label{K_intermediate} \,,
\end{align}
where
\begin{align}
	Q_{n,k}(u,u')
	&= 2 i \int_{-1}^1 d\nu \, \left( \frac{\partial}{\partial u} - \frac{\partial}{\partial u'} \right) \frac{\exp \left( -\frac{u^2}{4 [n - k - i (2 + \nu)]} - \frac{{u'}^2}{4 [k + i (2 + \nu)]} \right)}{\sqrt{[n - k - i (2 + \nu)] [k + i (2 + \nu)]}} \nonumber \\
	&= 2 i \int_1^3 dz \, \left( \frac{\partial}{\partial u} - \frac{\partial}{\partial u'} \right) \frac{\exp \left( -i \frac{u^2}{4 [z + i (n-k)]} + i \frac{{u'}^2}{4 (z - i k)} \right)}{\sqrt{[z + i (n-k)] (z - i k)}} \nonumber \\
	&= \int_1^3 dz \, \left( \frac{u}{z + i (n-k)} + \frac{u'}{z - i k} \right) \frac{\exp \left( -i \frac{u^2}{4 [z + i (n-k)]} + i \frac{{u'}^2}{4 (z - i k)} \right)}{\sqrt{[z + i (n-k)] (z - i k)}} \,. \label{Q_nk}
\end{align}
Finally, substituting Eq.~\eqref{Q_nk} into Eq.~\eqref{K_intermediate}, we arrive at Eq.~\eqref{K_expansion}

For $n=k=0$, the integral in Eq.~\eqref{Q_nk} can be evaluated as follows:
\begin{align*}
	Q_{0,0}(u,u')
	&= (u + u') \int_1^3 \frac{dz}{z^2} e^{-i \left( u^2 - {u'}^2 \right) / 4 z} = (u + u') \int_{1/3}^1 d\zeta \, e^{-i \left( u^2 - {u'}^2 \right) \zeta / 4} \\
	&= (u + u') \frac{e^{-i \left( u^2 - {u'}^2 \right) / 4} - e^{-i \left( u^2 - {u'}^2 \right) / 12}}{-i \left( u^2 - {u'}^2 \right)/4} = 8 e^{-i u^2 / 6} \frac{\sin \left[ \frac{1}{12} \left( u^2 - {u'}^2 \right) \right]}{u - u'} e^{i {u'}^2 / 6} \,.
\end{align*}

\twocolumngrid


%


\begin{thebibliography}{26}%
\makeatletter
\providecommand \@ifxundefined [1]{%
 \@ifx{#1\undefined}
}%
\providecommand \@ifnum [1]{%
 \ifnum #1\expandafter \@firstoftwo
 \else \expandafter \@secondoftwo
 \fi
}%
\providecommand \@ifx [1]{%
 \ifx #1\expandafter \@firstoftwo
 \else \expandafter \@secondoftwo
 \fi
}%
\providecommand \natexlab [1]{#1}%
\providecommand \enquote  [1]{``#1''}%
\providecommand \bibnamefont  [1]{#1}%
\providecommand \bibfnamefont [1]{#1}%
\providecommand \citenamefont [1]{#1}%
\providecommand \href@noop [0]{\@secondoftwo}%
\providecommand \href [0]{\begingroup \@sanitize@url \@href}%
\providecommand \@href[1]{\@@startlink{#1}\@@href}%
\providecommand \@@href[1]{\endgroup#1\@@endlink}%
\providecommand \@sanitize@url [0]{\catcode `\\12\catcode `\$12\catcode
  `\&12\catcode `\#12\catcode `\^12\catcode `\_12\catcode `\%12\relax}%
\providecommand \@@startlink[1]{}%
\providecommand \@@endlink[0]{}%
\providecommand \url  [0]{\begingroup\@sanitize@url \@url }%
\providecommand \@url [1]{\endgroup\@href {#1}{\urlprefix }}%
\providecommand \urlprefix  [0]{URL }%
\providecommand \Eprint [0]{\href }%
\providecommand \doibase [0]{http://dx.doi.org/}%
\providecommand \selectlanguage [0]{\@gobble}%
\providecommand \bibinfo  [0]{\@secondoftwo}%
\providecommand \bibfield  [0]{\@secondoftwo}%
\providecommand \translation [1]{[#1]}%
\providecommand \BibitemOpen [0]{}%
\providecommand \bibitemStop [0]{}%
\providecommand \bibitemNoStop [0]{.\EOS\space}%
\providecommand \EOS [0]{\spacefactor3000\relax}%
\providecommand \BibitemShut  [1]{\csname bibitem#1\endcsname}%
\let\auto@bib@innerbib\@empty
\bibitem [{\citenamefont {Allcock}(1969)}]{All69time-c}%
  \BibitemOpen
  \bibfield  {author} {\bibinfo {author} {\bibfnamefont {G.~R.}\ \bibnamefont
  {Allcock}},\ }\bibfield  {title} {\enquote {\bibinfo {title} {{The time of
  arrival in quantum mechanics III. The measurement ensemble}},}\ }\href
  {\doibase 10.1016/0003-4916(69)90253-X} {\bibfield  {journal} {\bibinfo
  {journal} {Ann. Phys. (N. Y).}\ }\textbf {\bibinfo {volume} {53}},\ \bibinfo
  {pages} {311} (\bibinfo {year} {1969})}\BibitemShut {NoStop}%
\bibitem [{\citenamefont {Kijowski}(1974)}]{Kij74time}%
  \BibitemOpen
  \bibfield  {author} {\bibinfo {author} {\bibfnamefont {J.}~\bibnamefont
  {Kijowski}},\ }\bibfield  {title} {\enquote {\bibinfo {title} {{On the time
  operator in quantum mechanics and the Heisenberg uncertainty relation for
  energy and time}},}\ }\href {\doibase 10.1016/S0034-4877(74)80004-2}
  {\bibfield  {journal} {\bibinfo  {journal} {Rep. Math. Phys.}\ }\textbf
  {\bibinfo {volume} {6}},\ \bibinfo {pages} {361} (\bibinfo {year}
  {1974})}\BibitemShut {NoStop}%
\bibitem [{\citenamefont {Bracken}\ and\ \citenamefont
  {Melloy}(1994)}]{BM94Probability}%
  \BibitemOpen
  \bibfield  {author} {\bibinfo {author} {\bibfnamefont {A.~J.}\ \bibnamefont
  {Bracken}}\ and\ \bibinfo {author} {\bibfnamefont {G.~F.}\ \bibnamefont
  {Melloy}},\ }\bibfield  {title} {\enquote {\bibinfo {title} {{Probability
  backflow and a new dimensionless quantum number}},}\ }\href {\doibase
  10.1088/0305-4470/27/6/040} {\bibfield  {journal} {\bibinfo  {journal} {J.
  Phys. A: Math. Gen.}\ }\textbf {\bibinfo {volume} {27}},\ \bibinfo {pages}
  {2197} (\bibinfo {year} {1994})}\BibitemShut {NoStop}%
\bibitem [{\citenamefont {Penz}\ \emph {et~al.}(2006)\citenamefont {Penz},
  \citenamefont {Gr{\"{u}}bl}, \citenamefont {Kreidl},\ and\ \citenamefont
  {Wagner}}]{PGKW06new}%
  \BibitemOpen
  \bibfield  {author} {\bibinfo {author} {\bibfnamefont {M.}~\bibnamefont
  {Penz}}, \bibinfo {author} {\bibfnamefont {G.}~\bibnamefont {Gr{\"{u}}bl}},
  \bibinfo {author} {\bibfnamefont {S.}~\bibnamefont {Kreidl}}, \ and\ \bibinfo
  {author} {\bibfnamefont {P.}~\bibnamefont {Wagner}},\ }\bibfield  {title}
  {\enquote {\bibinfo {title} {{A new approach to quantum backflow}},}\ }\href
  {\doibase 10.1088/0305-4470/39/2/012} {\bibfield  {journal} {\bibinfo
  {journal} {J. Phys. A: Math. Gen.}\ }\textbf {\bibinfo {volume} {39}},\
  \bibinfo {pages} {423} (\bibinfo {year} {2006})}\BibitemShut {NoStop}%
\bibitem [{\citenamefont {Bracken}\ and\ \citenamefont
  {Melloy}(2014)}]{BM14Waiting}%
  \BibitemOpen
  \bibfield  {author} {\bibinfo {author} {\bibfnamefont {A.~J.}\ \bibnamefont
  {Bracken}}\ and\ \bibinfo {author} {\bibfnamefont {G.~F.}\ \bibnamefont
  {Melloy}},\ }\bibfield  {title} {\enquote {\bibinfo {title} {{Waiting for the
  quantum bus: The flow of negative probability}},}\ }\href {\doibase
  10.1016/j.shpsb.2014.09.001} {\bibfield  {journal} {\bibinfo  {journal}
  {Stud. Hist. Philos. Sci. Part B Stud. Hist. Philos. Mod. Phys.}\ }\textbf
  {\bibinfo {volume} {48}},\ \bibinfo {pages} {13} (\bibinfo {year}
  {2014})}\BibitemShut {NoStop}%
\bibitem [{\citenamefont {Yearsley}\ \emph {et~al.}(2012)\citenamefont
  {Yearsley}, \citenamefont {Halliwell}, \citenamefont {Hartshorn},\ and\
  \citenamefont {Whitby}}]{YHHW12Analytical}%
  \BibitemOpen
  \bibfield  {author} {\bibinfo {author} {\bibfnamefont {J.~M.}\ \bibnamefont
  {Yearsley}}, \bibinfo {author} {\bibfnamefont {J.~J.}\ \bibnamefont
  {Halliwell}}, \bibinfo {author} {\bibfnamefont {R.}~\bibnamefont
  {Hartshorn}}, \ and\ \bibinfo {author} {\bibfnamefont {A.}~\bibnamefont
  {Whitby}},\ }\bibfield  {title} {\enquote {\bibinfo {title} {{Analytical
  examples, measurement models, and classical limit of quantum backflow}},}\
  }\href {\doibase 10.1103/PhysRevA.86.042116} {\bibfield  {journal} {\bibinfo
  {journal} {Phys. Rev. A}\ }\textbf {\bibinfo {volume} {86}},\ \bibinfo
  {pages} {042116} (\bibinfo {year} {2012})}\BibitemShut {NoStop}%
\bibitem [{\citenamefont {Eveson}\ \emph {et~al.}(2005)\citenamefont {Eveson},
  \citenamefont {Fewster},\ and\ \citenamefont {Verch}}]{EFV05Quantum}%
  \BibitemOpen
  \bibfield  {author} {\bibinfo {author} {\bibfnamefont {S.~P.}\ \bibnamefont
  {Eveson}}, \bibinfo {author} {\bibfnamefont {C.~J.}\ \bibnamefont {Fewster}},
  \ and\ \bibinfo {author} {\bibfnamefont {R.}~\bibnamefont {Verch}},\
  }\bibfield  {title} {\enquote {\bibinfo {title} {{Quantum Inequalities in
  Quantum Mechanics}},}\ }\href {\doibase 10.1007/s00023-005-0197-9} {\bibfield
   {journal} {\bibinfo  {journal} {Ann. Henri Poincar\'e}\ }\textbf {\bibinfo
  {volume} {6}},\ \bibinfo {pages} {1} (\bibinfo {year} {2005})}\BibitemShut
  {NoStop}%
\bibitem [{\citenamefont {Halliwell}\ \emph {et~al.}(2013)\citenamefont
  {Halliwell}, \citenamefont {Gillman}, \citenamefont {Lennon}, \citenamefont
  {Patel},\ and\ \citenamefont {Ramirez}}]{HGL+13Quantum}%
  \BibitemOpen
  \bibfield  {author} {\bibinfo {author} {\bibfnamefont {J.~J.}\ \bibnamefont
  {Halliwell}}, \bibinfo {author} {\bibfnamefont {E.}~\bibnamefont {Gillman}},
  \bibinfo {author} {\bibfnamefont {O.}~\bibnamefont {Lennon}}, \bibinfo
  {author} {\bibfnamefont {M.}~\bibnamefont {Patel}}, \ and\ \bibinfo {author}
  {\bibfnamefont {I.}~\bibnamefont {Ramirez}},\ }\bibfield  {title} {\enquote
  {\bibinfo {title} {{Quantum backflow states from eigenstates of the
  regularized current operator}},}\ }\href {\doibase
  10.1088/1751-8113/46/47/475303} {\bibfield  {journal} {\bibinfo  {journal}
  {J. Phys. A: Math. Theor.}\ }\textbf {\bibinfo {volume} {46}},\ \bibinfo
  {pages} {475303} (\bibinfo {year} {2013})}\BibitemShut {NoStop}%
\bibitem [{\citenamefont {Muga}\ \emph {et~al.}(1999)\citenamefont {Muga},
  \citenamefont {Palao},\ and\ \citenamefont {Leavens}}]{MPL99Arrival}%
  \BibitemOpen
  \bibfield  {author} {\bibinfo {author} {\bibfnamefont {J.~G.}\ \bibnamefont
  {Muga}}, \bibinfo {author} {\bibfnamefont {J.~P.}\ \bibnamefont {Palao}}, \
  and\ \bibinfo {author} {\bibfnamefont {C.~R.}\ \bibnamefont {Leavens}},\
  }\bibfield  {title} {\enquote {\bibinfo {title} {{Arrival time distributions
  and perfect absorption in classical and quantum mechanics}},}\ }\href
  {\doibase 10.1016/S0375-9601(99)00020-1} {\bibfield  {journal} {\bibinfo
  {journal} {Phys. Lett. A}\ }\textbf {\bibinfo {volume} {253}},\ \bibinfo
  {pages} {21} (\bibinfo {year} {1999})}\BibitemShut {NoStop}%
\bibitem [{\citenamefont {Muga}\ and\ \citenamefont
  {Leavens}(2000)}]{ML00Arrival}%
  \BibitemOpen
  \bibfield  {author} {\bibinfo {author} {\bibfnamefont {J.~G.}\ \bibnamefont
  {Muga}}\ and\ \bibinfo {author} {\bibfnamefont {C.~R.}\ \bibnamefont
  {Leavens}},\ }\bibfield  {title} {\enquote {\bibinfo {title} {{Arrival time
  in quantum mechanics}},}\ }\href {\doibase 10.1016/S0370-1573(00)00047-8}
  {\bibfield  {journal} {\bibinfo  {journal} {Phys. Rep.}\ }\textbf {\bibinfo
  {volume} {338}},\ \bibinfo {pages} {353} (\bibinfo {year}
  {2000})}\BibitemShut {NoStop}%
\bibitem [{\citenamefont {Halliwell}\ \emph {et~al.}(2019)\citenamefont
  {Halliwell}, \citenamefont {Beck}, \citenamefont {Lee},\ and\ \citenamefont
  {O'Brien}}]{HBLO19Quasiprobability}%
  \BibitemOpen
  \bibfield  {author} {\bibinfo {author} {\bibfnamefont {J.~J.}\ \bibnamefont
  {Halliwell}}, \bibinfo {author} {\bibfnamefont {H.}~\bibnamefont {Beck}},
  \bibinfo {author} {\bibfnamefont {B.~K.~B.}\ \bibnamefont {Lee}}, \ and\
  \bibinfo {author} {\bibfnamefont {S.}~\bibnamefont {O'Brien}},\ }\bibfield
  {title} {\enquote {\bibinfo {title} {{Quasiprobability for the arrival-time
  problem with links to backflow and the Leggett-Garg inequalities}},}\ }\href
  {\doibase 10.1103/PhysRevA.99.012124} {\bibfield  {journal} {\bibinfo
  {journal} {Phys. Rev. A}\ }\textbf {\bibinfo {volume} {99}},\ \bibinfo
  {pages} {012124} (\bibinfo {year} {2019})}\BibitemShut {NoStop}%
\bibitem [{\citenamefont {Berry}(2010)}]{Ber10Quantum}%
  \BibitemOpen
  \bibfield  {author} {\bibinfo {author} {\bibfnamefont {M.~V.}\ \bibnamefont
  {Berry}},\ }\bibfield  {title} {\enquote {\bibinfo {title} {{Quantum
  backflow, negative kinetic energy, and optical retro-propagation}},}\ }\href
  {\doibase 10.1088/1751-8113/43/41/415302} {\bibfield  {journal} {\bibinfo
  {journal} {J. Phys. A: Math. Theor.}\ }\textbf {\bibinfo {volume} {43}},\
  \bibinfo {pages} {415302} (\bibinfo {year} {2010})}\BibitemShut {NoStop}%
\bibitem [{\citenamefont {Bostelmann}\ \emph {et~al.}(2017)\citenamefont
  {Bostelmann}, \citenamefont {Cadamuro},\ and\ \citenamefont
  {Lechner}}]{BCL17Quantum}%
  \BibitemOpen
  \bibfield  {author} {\bibinfo {author} {\bibfnamefont {H.}~\bibnamefont
  {Bostelmann}}, \bibinfo {author} {\bibfnamefont {D.}~\bibnamefont
  {Cadamuro}}, \ and\ \bibinfo {author} {\bibfnamefont {G.}~\bibnamefont
  {Lechner}},\ }\bibfield  {title} {\enquote {\bibinfo {title} {{Quantum
  backflow and scattering}},}\ }\href {\doibase 10.1103/PhysRevA.96.012112}
  {\bibfield  {journal} {\bibinfo  {journal} {Phys. Rev. A}\ }\textbf {\bibinfo
  {volume} {96}},\ \bibinfo {pages} {012112} (\bibinfo {year}
  {2017})}\BibitemShut {NoStop}%
\bibitem [{\citenamefont {Strange}(2012)}]{Str12Large}%
  \BibitemOpen
  \bibfield  {author} {\bibinfo {author} {\bibfnamefont {P.}~\bibnamefont
  {Strange}},\ }\bibfield  {title} {\enquote {\bibinfo {title} {{Large quantum
  probability backflow and the azimuthal angle–angular momentum uncertainty
  relation for an electron in a constant magnetic field}},}\ }\href {\doibase
  10.1088/0143-0807/33/5/1147} {\bibfield  {journal} {\bibinfo  {journal} {Eur.
  J. Phys.}\ }\textbf {\bibinfo {volume} {33}},\ \bibinfo {pages} {1147}
  (\bibinfo {year} {2012})}\BibitemShut {NoStop}%
\bibitem [{\citenamefont {Palmero}\ \emph {et~al.}(2013)\citenamefont
  {Palmero}, \citenamefont {Torrontegui}, \citenamefont {Muga},\ and\
  \citenamefont {Modugno}}]{PTMM13Detecting}%
  \BibitemOpen
  \bibfield  {author} {\bibinfo {author} {\bibfnamefont {M.}~\bibnamefont
  {Palmero}}, \bibinfo {author} {\bibfnamefont {E.}~\bibnamefont
  {Torrontegui}}, \bibinfo {author} {\bibfnamefont {J.~G.}\ \bibnamefont
  {Muga}}, \ and\ \bibinfo {author} {\bibfnamefont {M.}~\bibnamefont
  {Modugno}},\ }\bibfield  {title} {\enquote {\bibinfo {title} {{Detecting
  quantum backflow by the density of a Bose-Einstein condensate}},}\ }\href
  {\doibase 10.1103/PhysRevA.87.053618} {\bibfield  {journal} {\bibinfo
  {journal} {Phys. Rev. A}\ }\textbf {\bibinfo {volume} {87}},\ \bibinfo
  {pages} {053618} (\bibinfo {year} {2013})}\BibitemShut {NoStop}%
\bibitem [{\citenamefont {Melloy}\ and\ \citenamefont
  {Bracken}(1998{\natexlab{a}})}]{MB98velocity}%
  \BibitemOpen
  \bibfield  {author} {\bibinfo {author} {\bibfnamefont {G.~F.}\ \bibnamefont
  {Melloy}}\ and\ \bibinfo {author} {\bibfnamefont {A.~J.}\ \bibnamefont
  {Bracken}},\ }\bibfield  {title} {\enquote {\bibinfo {title} {{The velocity
  of probability transport in quantum mechanics}},}\ }\href {\doibase
  10.1002/(SICI)1521-3889(199812)7:7/8<726::AID-ANDP726>3.0.CO;2-P} {\bibfield
  {journal} {\bibinfo  {journal} {Ann. Phys.}\ }\textbf {\bibinfo {volume}
  {7}},\ \bibinfo {pages} {726} (\bibinfo {year}
  {1998}{\natexlab{a}})}\BibitemShut {NoStop}%
\bibitem [{\citenamefont {Mardonov}\ \emph {et~al.}(2014)\citenamefont
  {Mardonov}, \citenamefont {Palmero}, \citenamefont {Modugno}, \citenamefont
  {Sherman},\ and\ \citenamefont {Muga}}]{MPM+14Interference}%
  \BibitemOpen
  \bibfield  {author} {\bibinfo {author} {\bibfnamefont {S.}~\bibnamefont
  {Mardonov}}, \bibinfo {author} {\bibfnamefont {M.}~\bibnamefont {Palmero}},
  \bibinfo {author} {\bibfnamefont {M.}~\bibnamefont {Modugno}}, \bibinfo
  {author} {\bibfnamefont {E.~Y.}\ \bibnamefont {Sherman}}, \ and\ \bibinfo
  {author} {\bibfnamefont {J.~G.}\ \bibnamefont {Muga}},\ }\bibfield  {title}
  {\enquote {\bibinfo {title} {{Interference of spin-orbit–coupled
  Bose-Einstein condensates}},}\ }\href {\doibase 10.1209/0295-5075/106/60004}
  {\bibfield  {journal} {\bibinfo  {journal} {EPL (Europhysics Lett.)}\
  }\textbf {\bibinfo {volume} {106}},\ \bibinfo {pages} {60004} (\bibinfo
  {year} {2014})}\BibitemShut {NoStop}%
\bibitem [{\citenamefont {Albarelli}\ \emph {et~al.}(2016)\citenamefont
  {Albarelli}, \citenamefont {Guaita},\ and\ \citenamefont
  {Paris}}]{AGP16Quantum}%
  \BibitemOpen
  \bibfield  {author} {\bibinfo {author} {\bibfnamefont {F.}~\bibnamefont
  {Albarelli}}, \bibinfo {author} {\bibfnamefont {T.}~\bibnamefont {Guaita}}, \
  and\ \bibinfo {author} {\bibfnamefont {M.~G.~A.}\ \bibnamefont {Paris}},\
  }\bibfield  {title} {\enquote {\bibinfo {title} {{Quantum backflow effect and
  nonclassicality}},}\ }\href {\doibase 10.1142/S0219749916500325} {\bibfield
  {journal} {\bibinfo  {journal} {Int. J. Quantum Inf.}\ }\textbf {\bibinfo
  {volume} {14}},\ \bibinfo {pages} {1650032} (\bibinfo {year}
  {2016})}\BibitemShut {NoStop}%
\bibitem [{\citenamefont {Mousavi}\ and\ \citenamefont
  {Miret-Art{\'{e}}s}(2020)}]{MM20Dissipative}%
  \BibitemOpen
  \bibfield  {author} {\bibinfo {author} {\bibfnamefont {S.~V.}\ \bibnamefont
  {Mousavi}}\ and\ \bibinfo {author} {\bibfnamefont {S.}~\bibnamefont
  {Miret-Art{\'{e}}s}},\ }\bibfield  {title} {\enquote {\bibinfo {title}
  {{Dissipative quantum backflow}},}\ }\href {\doibase
  10.1140/epjp/s13360-020-00336-5} {\bibfield  {journal} {\bibinfo  {journal}
  {Eur. Phys. J. Plus}\ }\textbf {\bibinfo {volume} {135}},\ \bibinfo {pages}
  {324} (\bibinfo {year} {2020})}\BibitemShut {NoStop}%
\bibitem [{\citenamefont {Melloy}\ and\ \citenamefont
  {Bracken}(1998{\natexlab{b}})}]{MB98Probability}%
  \BibitemOpen
  \bibfield  {author} {\bibinfo {author} {\bibfnamefont {G.~F.}\ \bibnamefont
  {Melloy}}\ and\ \bibinfo {author} {\bibfnamefont {A.~J.}\ \bibnamefont
  {Bracken}},\ }\bibfield  {title} {\enquote {\bibinfo {title} {{Probability
  Backflow for a Dirac Particle}},}\ }\href {\doibase 10.1023/A:1018724313788}
  {\bibfield  {journal} {\bibinfo  {journal} {Found. Phys.}\ }\textbf {\bibinfo
  {volume} {28}},\ \bibinfo {pages} {505} (\bibinfo {year}
  {1998}{\natexlab{b}})}\BibitemShut {NoStop}%
\bibitem [{\citenamefont {Su}\ and\ \citenamefont {Chen}(2018)}]{SC18Quantum}%
  \BibitemOpen
  \bibfield  {author} {\bibinfo {author} {\bibfnamefont {H.}~\bibnamefont
  {Su}}\ and\ \bibinfo {author} {\bibfnamefont {J.}~\bibnamefont {Chen}},\
  }\bibfield  {title} {\enquote {\bibinfo {title} {{Quantum backflow in
  solutions to the Dirac equation of the spin-1/2 free particle}},}\ }\href
  {\doibase 10.1142/S0217732318501869} {\bibfield  {journal} {\bibinfo
  {journal} {Mod. Phys. Lett. A}\ }\textbf {\bibinfo {volume} {33}},\ \bibinfo
  {pages} {1850186} (\bibinfo {year} {2018})}\BibitemShut {NoStop}%
\bibitem [{\citenamefont {Ashfaque}\ \emph {et~al.}(2019)\citenamefont
  {Ashfaque}, \citenamefont {Lynch},\ and\ \citenamefont
  {Strange}}]{ALS19Relativistic}%
  \BibitemOpen
  \bibfield  {author} {\bibinfo {author} {\bibfnamefont {J.~M.}\ \bibnamefont
  {Ashfaque}}, \bibinfo {author} {\bibfnamefont {J.}~\bibnamefont {Lynch}}, \
  and\ \bibinfo {author} {\bibfnamefont {P.}~\bibnamefont {Strange}},\
  }\bibfield  {title} {\enquote {\bibinfo {title} {{Relativistic quantum
  backflow}},}\ }\href {\doibase 10.1088/1402-4896/ab265c} {\bibfield
  {journal} {\bibinfo  {journal} {Phys. Scr.}\ }\textbf {\bibinfo {volume}
  {94}},\ \bibinfo {pages} {125107} (\bibinfo {year} {2019})}\BibitemShut
  {NoStop}%
\bibitem [{\citenamefont {Eliezer}\ \emph {et~al.}(2020)\citenamefont
  {Eliezer}, \citenamefont {Zacharias},\ and\ \citenamefont
  {Bahabad}}]{EZB20Observation}%
  \BibitemOpen
  \bibfield  {author} {\bibinfo {author} {\bibfnamefont {Y.}~\bibnamefont
  {Eliezer}}, \bibinfo {author} {\bibfnamefont {T.}~\bibnamefont {Zacharias}},
  \ and\ \bibinfo {author} {\bibfnamefont {A.}~\bibnamefont {Bahabad}},\
  }\bibfield  {title} {\enquote {\bibinfo {title} {{Observation of optical
  backflow}},}\ }\href {\doibase 10.1364/OPTICA.371494} {\bibfield  {journal}
  {\bibinfo  {journal} {Optica}\ }\textbf {\bibinfo {volume} {7}},\ \bibinfo
  {pages} {72} (\bibinfo {year} {2020})}\BibitemShut {NoStop}%
\bibitem [{\citenamefont {Goussev}(2019)}]{Gou19Equivalence}%
  \BibitemOpen
  \bibfield  {author} {\bibinfo {author} {\bibfnamefont {A.}~\bibnamefont
  {Goussev}},\ }\bibfield  {title} {\enquote {\bibinfo {title} {{Equivalence
  between quantum backflow and classically forbidden probability flow in a
  diffraction-in-time problem}},}\ }\href {\doibase 10.1103/PhysRevA.99.043626}
  {\bibfield  {journal} {\bibinfo  {journal} {Phys. Rev. A}\ }\textbf {\bibinfo
  {volume} {99}},\ \bibinfo {pages} {043626} (\bibinfo {year}
  {2019})}\BibitemShut {NoStop}%
\bibitem [{\citenamefont {van Dijk}\ and\ \citenamefont
  {Toyama}(2019)}]{DT19Decay}%
  \BibitemOpen
  \bibfield  {author} {\bibinfo {author} {\bibfnamefont {W.}~\bibnamefont {van
  Dijk}}\ and\ \bibinfo {author} {\bibfnamefont {F.~M.}\ \bibnamefont
  {Toyama}},\ }\bibfield  {title} {\enquote {\bibinfo {title} {{Decay of a
  quasistable quantum system and quantum backflow}},}\ }\href {\doibase
  10.1103/PhysRevA.100.052101} {\bibfield  {journal} {\bibinfo  {journal}
  {Phys. Rev. A}\ }\textbf {\bibinfo {volume} {100}},\ \bibinfo {pages}
  {052101} (\bibinfo {year} {2019})}\BibitemShut {NoStop}%
\bibitem [{Note1()}]{Note1}%
  \BibitemOpen
  \bibinfo {note} {In general, the sum of two unbounded Hermitian operators is
  not necessarily Hermitian. However, the operator $S(\protect \mathaccentV
  {hat}05E{p}) - \protect \mathaccentV {hat}05E{x}$ is a unitary transformation
  of the operator $-\protect \mathaccentV {hat}05E{x}$, and as such is
  Hermitian. Indeed, it is straightforward to check that $S(\protect
  \mathaccentV {hat}05E{p}) - \protect \mathaccentV {hat}05E{x} = \protect
  \mathaccentV {hat}05E{U}^{-1} (-\protect \mathaccentV {hat}05E{x}) \protect
  \mathaccentV {hat}05E{U}$, with $\protect \mathaccentV {hat}05E{U} = e^{i
  R(\protect \mathaccentV {hat}05E{p}) / \hbar }$ and $R(p) = \DOTSI \intop
  \ilimits@ ^{p} dq \protect \tmspace +\thinmuskip {.1667em} S(q)$
  (cf.~Eq.~\protect \textup {\hbox {\mathsurround \z@ \protect \normalfont
  (\ignorespaces \ref {w_mom_rep}\unskip \@@italiccorr )}})}\BibitemShut
  {NoStop}%
\end{thebibliography}
\end{document}